\documentclass[12pt]{article}

\renewcommand{\normalsize}{%
  \@setfontsize\large\@xipt\@xiipt}

\usepackage{amsmath}
\usepackage{arxiv}
\usepackage[sort&compress, numbers]{natbib}
\usepackage{float}
\usepackage{array}
\usepackage{caption}
\usepackage{multirow}

\usepackage[utf8]{inputenc} 
\usepackage[T1]{fontenc}    
\usepackage{url}            
\usepackage{booktabs}       
\usepackage{amsfonts}       
\usepackage{nicefrac}       
\usepackage{microtype}      
\usepackage{lipsum}
\usepackage{graphicx}
\graphicspath{ {./images/} }

\title{Measuring and identifying factors of individuals' trust in Large Language Models}

\author{
 Edoardo Sebastiano De Duro \\
  Department of Psychology and Cognitive Science \\
  University of Trento, Italy \\
  \texttt{edoardo.deduro@unitn.it} \\
 \And
 Giuseppe Alessandro Veltri \\
  NUS Yong Loo Lin School of Medicine \\
  National University of Singapore, Singapore\\
  \texttt{gaveltri@nus.edu.sg} \\
\AND
 Hudson Golino \\
  Department of Psychology \\
  University of Virginia, USA \\
  \texttt{hfg9s@virginia.edu} \\
 \AND
 Massimo Stella \\
  Department of Psychology and Cognitive Science \\
  University of Trento, Italy \\
  \texttt{massimo.stella-1@unitn.it} \\
\AND
\\
}

\begin{document}

\maketitle

\begin{abstract}

Large Language Models (LLMs) can engage in human-looking conversational exchanges. Although conversations can elicit trust between users and LLMs, scarce empirical research has examined trust formation in human–LLM contexts, beyond LLMs' trustworthiness or human trust in AI in general. Here, we introduce the Trust-In-LLMs Index (TILLMI) as a new framework to measure individuals’ trust in LLMs, extending McAllister’s cognitive and affective trust dimensions to LLM-human interactions. We developed TILLMI as a psychometric scale, prototyped with a novel protocol we called LLM-simulated validity. The LLM-based scale was then validated in a sample of 1,000 US respondents. Exploratory Factor Analysis identified a two-factor structure. Two items were then removed due to redundancy, yielding a final 6-item scale with a 2-factor structure. Confirmatory Factor Analysis on a separate subsample showed strong model fit ($CFI = .995$, $TLI = .991$, $RMSEA = .046$, $p_{X^2} > .05$). Convergent validity analysis revealed that trust in LLMs correlated positively with openness to experience, extraversion, and cognitive flexibility, but negatively with neuroticism. Based on these findings, we interpreted TILLMI's factors as “closeness with LLMs” (affective dimension) and “reliance on LLMs” (cognitive dimension). Younger males exhibited higher closeness with- and reliance on LLMs compared to older women. Individuals with no direct experience with LLMs exhibited lower levels of trust compared to LLMs' users. These findings offer a novel empirical foundation for measuring trust in AI-driven verbal communication, informing responsible design, and fostering balanced human–AI collaboration.

\end{abstract}

\newpage

\section{Introduction}
Trust is a key factor shaping interactions in human settings. Trust fosters cooperation, reduces uncertainty, and supports the smooth flow of information \citep{McAllister1995, Mayer1995}. In psychology, trust has been studied for decades as a core determinant of team performance and stakeholder relationships. McAllister \citep{McAllister1995} described trust as both affective and cognitive. While affective trust involves emotional bonds and genuine care for others, cognitive trust relies on reasoned assessments of another party’s competence and reliability. Subsequent studies have shown how, despite being highly correlated, cognitive and affective trust are indeed empirically distinguishable \citep{johnson2005cognitive}. In workplaces and organizations, these dimensions help coworkers feel confident in each other's intentions and abilities. Employees who perceive both affective and cognitive trust in their teams often report higher job satisfaction, stronger commitment, and better overall performance \citep{Mayer1995}.

However, workplaces are rapidly changing. Many tasks once reserved for human employees are now assigned to automated tools and artificial intelligence (AI). Recent advances in AI, especially in Large Language Models (LLMs), have made machines more capable of handling tasks that require advanced reasoning and context-driven analysis \citep{breum2024persuasive,binz2025should,rossetti2024social}. Although LLMs evidently differ from human colleagues \cite{de2025introducing}, trust remains vital for effective collaboration \citep{Lee2004,liu2023trustworthy}. Without trust, users may hesitate to follow AI-generated advice, share sensitive information, or integrate these tools into daily workflows \cite{rossetti2024social}. Conversely, excessive trust may lead to undue reliance on systems that can still produce errors or biased responses \citep{binz2025should, dillion2023can, stella2023using, abramski2023cognitive}.

Researchers have long been concerned with how people form trust in technology. Madsen and Gregor \citep{Madsen2000} argued that system reliability, interface design, and user perceptions of risk play large roles in human–computer trust. McKnight and colleagues \citep{mcknight2011trust} offered a framework for measuring trust in specific technologies, highlighting factors such as structural assurances, situational normality, and a user’s general propensity to trust. Hoff and Bashir \citep{Hoff2015} expanded on this foundation by analysing how transparency, feedback, and system predictability contribute to trust in automation. In a meta-analysis, Schaefer and colleagues \citep{Schaefer2016} found that users rely on cues such as perceived reliability, ease of use, and contextual factors when judging whether to trust automated agents. Moreover, in an experimental setting \citep{gompei2018factors}, interaction with social robots underlined that the manipulation of different variables impacted differently affective and cognitive trust (e.g. topic of the conversation influenced affective trust while robot's mistakes shaped the cognitive one).

Our approach differs from past works testing the trustworthiness of the content produced by LLMs \citep{liu2023trustworthy,bo2024rely} and also by past works focusing on trust formation in various technological contexts \citep{Schaefer2016,gompei2018factors}. We argue that LLMs present unique challenges \citep{stella2023using} that extend beyond traditional human-computer trust dynamics. Unlike simpler automated systems or social robots, LLMs present the unique capability of engaging in natural language exchanges that closely mirror human ones per form, syntax and emotional tone \citep{de2025introducing,breum2024persuasive}. Thus, measuring trust in LLMs requires a dedicated psychometric scale. Prior scales, designed for trust in automation or simpler forms of AI \cite{sindermann2021assessing}, cannot fully capture the nuances of human–LLM interactions. Researchers have created instruments to measure user trust in various technological contexts, such as e-commerce, recommendation systems, online services, and autonomous systems \citep{mcknight2011trust, Gefen2003, Schaefer2016}. Others have focused on how design features, such as system transparency or perceived anthropomorphism, shape users’ willingness to rely on AI \citep{Bartneck2009, Hoff2015, Glikson2020}. Yet, large-scale text generation tools pose distinct challenges because they can produce fluent but potentially misleading or biased text. LLMs often function as “black boxes”, making it difficult for users to assess their decision processes \citep{Ribeiro2016, Jacovi2021, rossetti2024social}. 

A new instrument must integrate established ideas about affective and cognitive trust \citep{McAllister1995}, while also reflecting the affordances and risks specific to LLMs. On the one hand, affective trust describes the emotional bond between a user and a system, capturing whether the user feels assured or supported. On the other hand, cognitive trust is based on rational judgments of competence, reliability, and consistency. In the case of LLMs, these judgments hinge on factors like output accuracy, fairness, and the clarity of explanations provided by the system. Drawing from both components, a robust “trust equation” \citep{maister2021trusted} would thus need to weigh users’ emotional comfort (including aspects such as perceived benevolence) against their logical assessments of system performance \citep{Schoorman2007}.

Such a psychometric tool could help researchers pinpoint which factors—such as transparency, perceived competence, or ease of interaction—most strongly affect how users decide to trust or distrust an LLM. For instance, perceived competence or reliability may hinge on how often an LLM provides accurate, unbiased, or context-aware information. The ease of interaction, or closeness, could depend on the design of the user interface or personalization features that adapt to the needs of the user \citep{Madsen2000}. When these components merge, they shape whether a user sees the system as credible, dependable, and safe to rely upon.

Once available, a well-designed psychometric measure could guide developers on how to optimise LLMs or their feedback loops for further improving human-to-LLMs' trustworthiness. For example, if scale results reveal that perceived risk undermines trust in certain work settings, system designers might implement clearer disclaimers or provide confidence scores that reflect the LLM’s uncertainty \citep{Lee2004, Schaefer2016}. If cognitive trust emerges as a stronger predictor of LLM adoption in an organization, developers could focus on verifiable performance metrics or robust error handling procedures. In contrast, if affective trust proves important in settings where users feel anxious about new technologies, an LLM might include more empathetic language or user-centric design elements \citep{de2025introducing, Glikson2020}.


Finally, ethical concerns reinforce the need to study human trust in LLMs rather than LLMs' trustworthiness when producing content. When users trust large language models, the former accept vulnerability by disclosing personal information \cite{de2025introducing} or following automated decisions \citep{Glikson2020}. Undue vulnerability could create opportunities for harmful behavior, such as spreading misinformation or invading privacy \citep{breum2024persuasive}. Thus, measuring trust is essential for both leveraging the benefits of AI and guarding against its potential harms.

In this paper, we present the Trust-In-LLMs Index (TILLMI) as a new framework to measure individuals’ trust in LLMs. Grounded in McAllister’s \citep{McAllister1995} distinction between affective and cognitive trust, it also builds on research on human–machine trust \citep{Lee2004, Madsen2000, mcknight2011trust, Hoff2015, Schaefer2016}. We describe each step of the scale development process, including item generation, pilot testing, and validation. Our goal is to create a reliable and valid instrument that can guide both researchers and practitioners in understanding and shaping trust in LLMs. We are confident this work contributes to the broader discussion on how to integrate LLM-based systems into professional and everyday contexts in ways that are beneficial for science \citep{binz2025should}, transparent, and ethically sound.

\section{Results}

\subsection{Scale design}
The initial version of the Trust-In-LLM-Index featured 8 items (see \textit{Materials and Methods}). We outlined these items by drawing from McAllister's \citep{McAllister1995} conceptualisation of trust emerging in cooperative settings, such as workplaces and organizations. We combined that theoretical framework with relevant literature about the trust equation \cite{maister2021trusted} where trust is considered as being constituted by multiple intertwined components (credibility, reliability, intimacy and self-orientation). Half of the items of the TILLMI aimed at capturing the affective dimension of trust, while the remaining half focused on cognitive trust in LLMs. Throughout the scale design, we framed LLMs as tools rather than companions, inspired by recent approaches \citep{binz2025should}. Our novel psychometric scale employed a 5-level Likert-type scale with scores ranging from 1 (strongly disagree) to 5 (strongly agree). Following the initial scale design, a crucial step in scale development is the evaluation of item quality. To do so we employed a novel technique leveraging LLMs. We discuss works that adopted a similar approach and our implementation of this technique in the following sections.

\subsection{Motivating LLM-simulated validity}
Recent studies have shown that language embeddings and language models can effectively estimate psychometric measurements that traditionally relied on empirical data collection \cite{russell2024generative, dillion2023can}. In this way, LLMs can be used as (a) a way to simulate human participants and obtain large datasets of synthetic responses \cite{de2025introducing} and (b) tools to generate and evaluate items allowing to reduce the load on experts that are usually employed to assess items quality \cite{russell2024generative}. Inspired by these approaches combining LLMs and psychometrics, we conducted our own simulation test using LLMs as participants (see \textit{Materials and Methods}) to assess the designed items' quality, non-redundancy, and internal structure. We defined this approach as “LLM-simulated validity''. 

\subsection{LLM-simulated validity}
We conducted an Exploratory Factor Analysis (EFA) on the synthetic data. We computed the Kaiser-Meyer-Olkin (KMO), a measure of the proportion of variance that might be common among items. This index is useful to assess whether the data is suited for subsequent factor analysis. The results ($KMO = .86$) were above the commonly accepted threshold of $.80$ \cite{kaiser1974little}, hence we proceeded further with the EFA. To understand the number of factors emerging we used 2 methods. 
The first one (Kaiser method) identifies factors with eigenvalues greater than 1, as these indicate that a factor explains enough variance and is worth keeping. The second approach (parallel analysis; \cite{horn1965rationale}) is similar but compares each factor’s eigenvalues with generated ones from a Monte Carlo simulation, which randomises the data while preserving the dimension of the sample. The first method yielded a 2-factor solution, while the parallel analysis simulation approach suggested the presence of 3 distinct factors. For both solutions, the emerged factors yielded good internal consistency ($\alpha > .85$), meaning that the items belonging to each factor tended to measure related concepts.
Given the promising results of the simulated experiment, the questionnaire was thus administered to a total ($N = 1000$) of human participants. We collected the responses (see \textit{Materials and Methods}) and compared the results with those simulated with GPT-4.

\subsection{Comparison between LLMs and humans}
\begin{figure}
    \centering
    \includegraphics[width=0.8\linewidth]{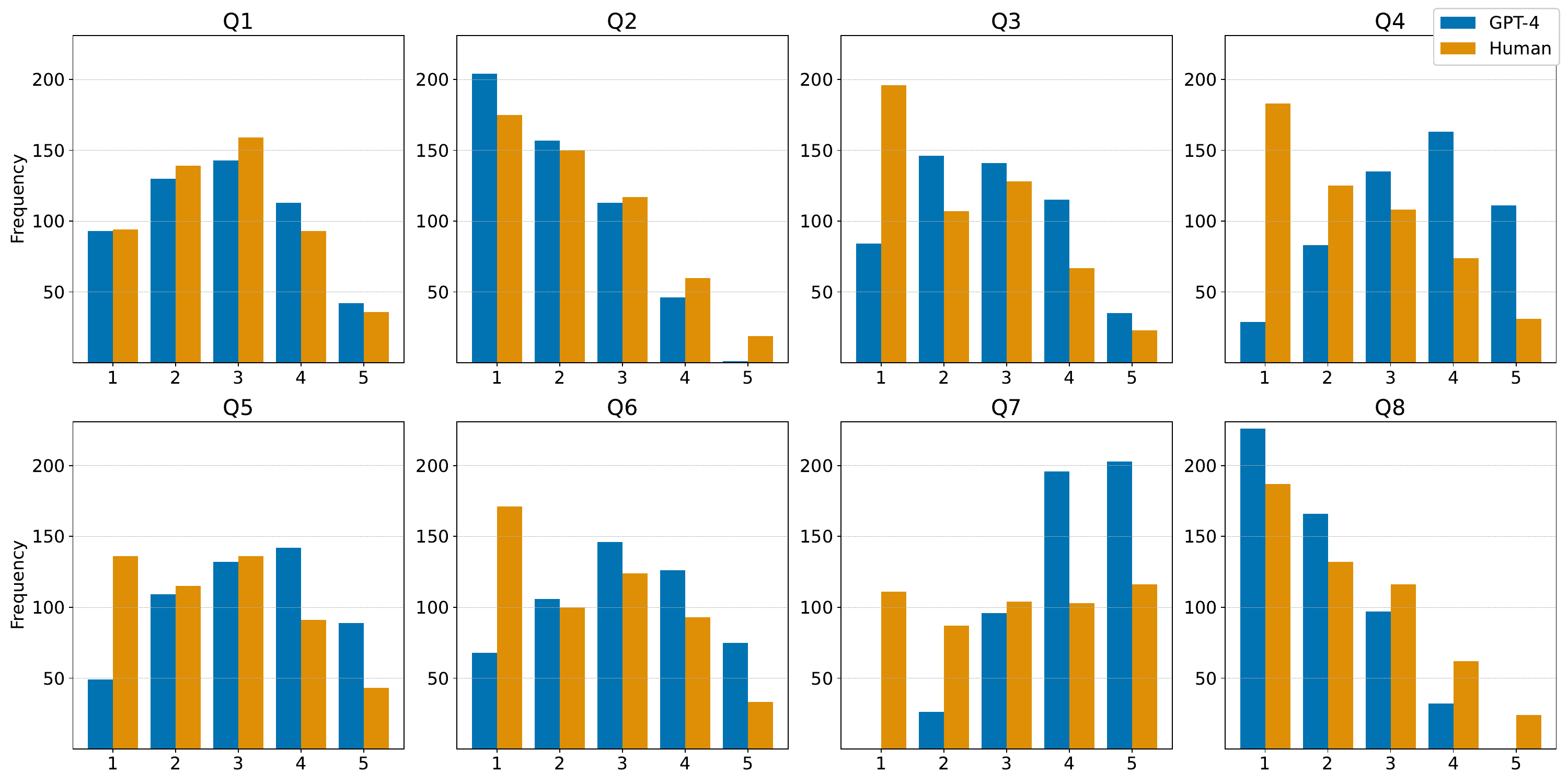}
    \caption{Response frequencies for 8 items of the initial TILLMI for  GPT-4 (in blue) and humans (in orange). To balance the 2 dataset ($n_{humans} = 521$, $n_{gpt4} = 800$) we extract a random sample of $n_{1} = 521$ from the synthetic GPT-4 dataset.}
    \label{fig:gpthuman}
\end{figure}

Figure \ref{fig:gpthuman} shows the comparison between GPT and human scores to the initial version of the TILLMI. Interestingly, some of the patterns found in humans were reproduced by GPT-4, suggesting the capability of LLMs to grasp the underlying psychological constructs relative to trust, in a way similar to humans. However, some differences emerged. The scoring pattern for Item 7 appears unusual compared to other items. In our scale higher scores usually mean greater trust in LLMs. However, Item 7 presents an inverse trend. A high score on its first part (“Despite trusting LLMs' results overall'') suggests high trust, but the full item (“Despite trusting LLMs' results overall, the last word is always mine'') implies low trust. It is possible that people's responses were based only on Item 7's opening clause rather than the complete item, while GPT-4 might have considered the whole item description. For items that were more clearly worded (e.g. Item 8), the scoring pattern between humans and LLMs is more aligned.



Following this comparison of score distributions between humans and LLMs, we conducted an Exploratory Graph Analysis (EGA) to evaluate the structure of our new tool.



\subsection{Exploratory Graph Analysis}
We carried out an Exploratory Graph Analysis (EGA; \cite{golino2017exploratory}) to identify the underlying structure of human participants' responses on the TILLMI. 
EGA can infer the empirical number of factors in complex psychometric datasets by leveraging network analysis to identify communities from items' correlations \citep{christensen2019estimating}. The following analyses include only respondents who had used LLMs at least once ($n_{1} = 521$).

The first step of the EGA involved performing the Unique Variable Analysis (UVA; \cite{christensen2023unique}), a technique useful to evaluate the local dependency of pairs of items (i.e., item redundancy) through weighted topological overlap (wTO; \cite{nowick2009differences}). Items with high wTO ($wTO > .3$) could, potentially, be removed without altering the internal structure of the network. In our case, none of the items showed high redundancy, so all 8-items were kept for further analysis.

\begin{figure*}
    \centering
    \includegraphics[width=0.65\linewidth]{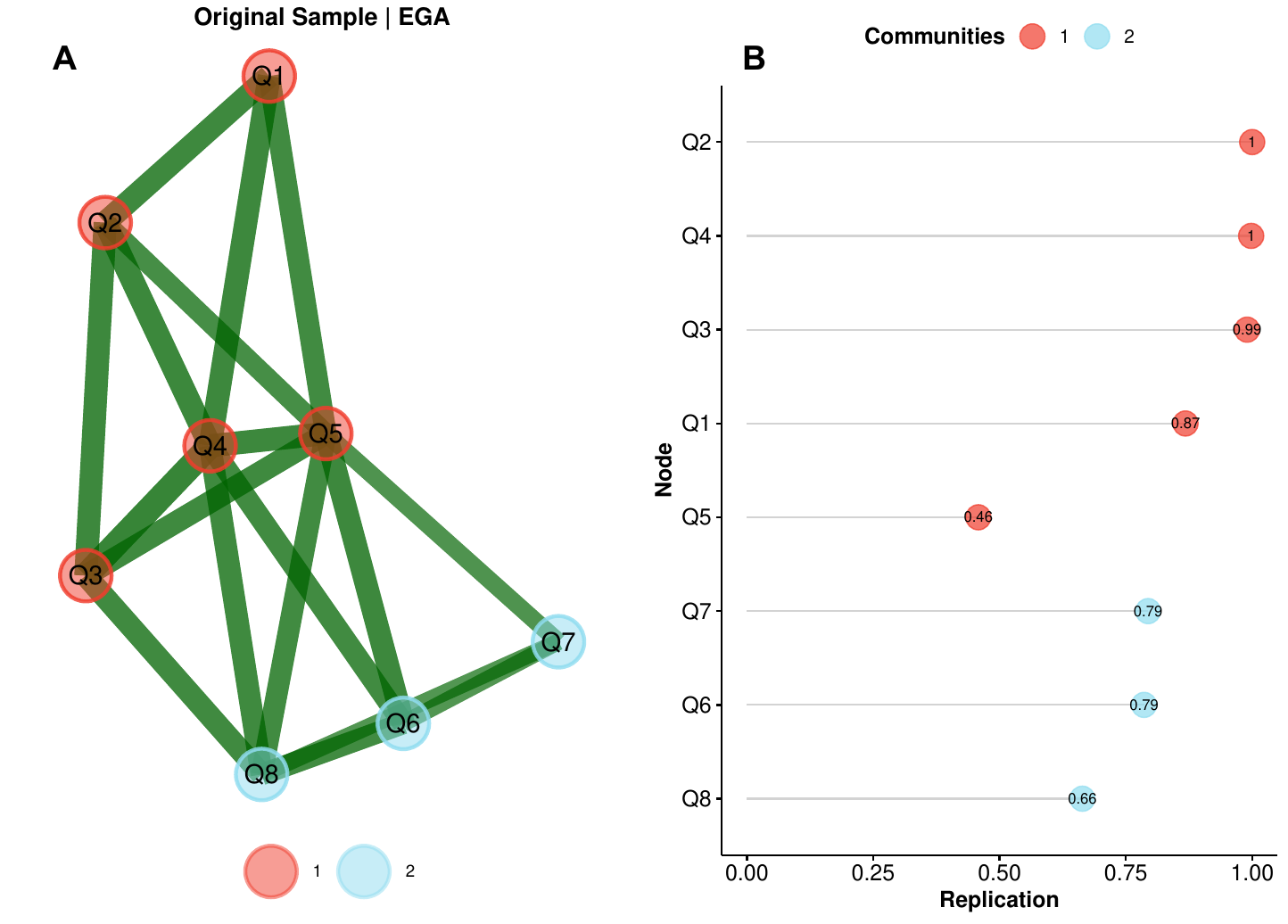}
    \caption{Exploratory Graph Analysis of the responses to the TILLMI for participants who stated to have used LLMs at least once ($n_{1} = 521$). (A) Psychometric network plotted using EGAnet. Nodes represent items of the TILLMI. Edges indicate the interaction between nodes, with green links representing positive interactions. (B) Item stability plot for the TILLMI bootstrap analysis. Probability of each item being assigned to its original dimension across bootstrap iterations is shown. Higher values (closer to 1) indicate that an item consistently appears in the same dimension, suggesting greater stability.}
    \label{fig:egafull}
\end{figure*}

Subsequently, we performed the EGA on the 8-item TILLMI scale. For further details on the parameters used, refer to the \textit{Materials and Methods} section. In the resulting psychometric networks, variables are represented as nodes, with their relationships visualised as edges. We show the network plot of the TILLMI in Figure \ref{fig:egafull}A. Overall, EGA suggested that a 2-factor structure fitted the data well. The correlations between items were positive, meaning that they tended to measure a similar construct. This was expected, as all the items were designed to capture trust in LLMs, where higher scores coded for higher trust. Nevertheless, our analysis revealed two distinct factors, highlighting that two facets of the same construct were captured by our novel scale.

Lastly, we performed an EGA bootstrap analysis (see \textit{Materials and Methods}), which involves repeatedly resampling the data and reestimating the network. In this way, we evaluated the stability and robustness of the estimated network structure. Two measures were taken into account: (a) Total Entropy Fit Index (TEFI; \cite{golino2021entropy}) and (b) item stability \cite{christensen2020psychometric, christensen2019estimating}. 

TEFI is a measure of the goodness of fit of the data to a model where lower values suggest a better fit. We compared the TEFI results of the bootstrap of our model, against a random 2-dimensional structure. The value found for the empirical 2-factor model ($TEFI = -3.7652, SD = 0.0267$) was significantly lower ($p = .006$) compared to a random 2-dimensional structure ($TEFI = -3.6682, SD = 0.0226$). Hence, the model emerging from human responses is better than a random model, suggesting the presence of an underlying structure in the data.

Item stability measures how consistently a variable appears in its originally estimated EGA dimension across repeated samplings. By doing so, it is possible to understand which nodes of a network show more stability. The results of the item stability analysis are shown in Figure \ref{fig:egafull}B. Certain items showed lower stability (Q1, Q5, Q6, Q7 and Q8) compared to the others (Q2, Q3, Q4), meaning that they could be not unique to a specific dimension. There are many reasons why this could be the case. For example, it might be possible that the EGA, while still identifying 2 factors, does not capture nuanced differences in two closely related facets of trust, i.e. in the considered sample size and with a limited number of items, the considered dimensions look very similar to each other. To gain further insights, we carried out a traditional Exploratory Factor Analysis to determine whether similar results were to be found.

\subsection{Exploratory Factor Analysis suitability}


To further examine the underlying structure of our data, we conducted an Exploratory Factor Analysis (EFA) on a subsample of the dataset (see \textit{Materials and Methods}) that was not used for the subsequent validation of the model in the Confirmatory Factor Analysis  (CFA). We begin with an assessment of the data's factorial suitability. To do so, we computed (a) the Kaiser-Meyer-Olkin index (KMO; \cite{kaiser1974little}) and (b) Bartlett's test of sphericity \cite{bartlett1951effect}. We already described the first method previously. Bartlett's test of sphericity is used to ensure that the correlation matrix between items differs from an identity matrix, where a significant result ($p < .05$) indicates that sufficient correlations exist among the variables to make factor analysis meaningful. In our case the results of the KMO ($KMO = .92$) and Bartlett's test ($\chi^2 = 1374.148$, $df = 28$, $p < .001$) suggested a significant degree of correlation between items, indicating that our sample is suitable for further analysis.

In addition, we checked whether the assumption of normality at a univariate level (Shapiro-Wilk test; \cite{shapiro1965analysis}) and multivariate level (Mardia test; \cite{mardia1970measures}) was met. In both cases, our data did not show normality ($W = .85572, p < .05 $; $z-kurtosis = 13.7, p < .05$). Given the non-normality of our data, we proceeded with the EFA and adopted principal axis factoring using the standard oblimin rotation method to obtain the initial results of our factor analysis.

\subsection{Selection of factor number for Exploratory Factor Analysis}
Having confirmed the data's suitability for factor analysis, we proceeded to determine the optimal number of factors using parallel analysis \cite{horn1965rationale} to compare the eigenvalues obtained from the test sample, against a model with uncorrelated variables. The number of factors resulting was equal to 2. Through subsequent analysis, we considered also the criterion of Very Simple Structure (VSS; \cite{revelle1979very}) and Velicer's Minimum Average Partial (MAP; \cite{velicer1976determining}) to further confirm the number of factors. Although some of the results hinted at a single-factor structure (Table \ref{tab:fit_vss}), the RMSEA index of fit revealed a lower fit for such a solution \cite{browne1992alternative, fabrigar1999evaluating} compared to the 2-factor model. How to address this apparent conflict in the data? Overall, the results suggest that a clear optimal number of factors is not evident from the data alone. As suggested in previous research \cite{devellis2021scale, beavers2019practical}, factor retention can be partially informed by theoretical insights and domain knowledge to determine the factor structure that is not only more “correct'', but also more meaningful. For this reasons, in light of (a) the previous distinction of trust in two facets (cognitive and affective; \cite{morrow2004cognitive, lewis1985, johnson2005cognitive}), (b) the rationale behind the design of the initial 8-items \cite{McAllister1995} we opted for the 2-factor model and continued with the Exploratory Factor Analysis.

\begin{table}
\centering
\fontsize{9}{10.5}\selectfont
\caption{Fit indices for 1-factor and 2-factor models. In bold, the best results are highlighted.}
\begin{tabular}{lccc}
\toprule
\textbf{Model} & \textbf{VSS} & \textbf{MAP} & \textbf{RMSEA} \\
\midrule
1-Factor  & \textbf{.95}  & \textbf{.039}  & .115 \\
2-Factor & .52 & .067 & \textbf{.077} \\
\bottomrule
\end{tabular}
\label{tab:fit_vss}
\end{table}

\subsection{Exploratory Factor Analysis}
Having selected the number of optimal factors, we ran the EFA using the parameters selected previously (\textit{principal axis factoring} as estimation method and \textit{oblimin} as rotation method). 
To assess the goodness of the EFA we consider (a) factor loadings (the level of the strength of the relation between factors and the observed variables), (b) complexities (representing the degree to which an item is unique to one factor) and (c) factor correlations (correlation between whole factors combined). The results of the first EFA highlighted the presence of some poor-quality items. Table \ref{tab:EFA} shows the results obtained. Items 1, 5 and 8 presented moderate levels of cross-loadings (items are associated with both factors exceeding our arbitrarily set threshold of $> |.3|$), indicating they did not exclusively associate with a single factor. All of these 3 items already showed low stability in the previous EGA. This is confirmed by the values of the complexities that showcased values diverging from 1. Nevertheless, factor correlations were good ($< 0.85$) \cite{brown2015confirmatory} confirming that it was worth keeping both the 2 factors.

Given the results found, we decided to drop Item 5 first and subsequently Item 1 (following the same procedure and rationale). The results of the second step of the EFA can be found in the \textit{SI Appendix, Table S1}. Finally, we ended up with a 6-item structure from the initial 8 items. In this model, Factor 1 accounts for 39.2\% of the total variance of the dataset, while Factor 2 explains 27.1\%. The results of the final EFA are presented in Table \ref{tab:efa_results}. Following the refinement to a 6-item model, we employed Confirmatory Factor Analysis to rigorously validate the structural validity of the TILLMI.


\begin{table}
\centering
\fontsize{9}{10.5}\selectfont
\caption{Factor loadings and component correlation for the 2-factor model. Cut thresholds for factor loadings are set to $|.3|$.}
\begin{tabular}{p{1.5cm} >{\centering\arraybackslash}p{1.2cm} >{\centering\arraybackslash}p{1.2cm} >{\centering\arraybackslash}p{1.2cm}}
\toprule
\textbf{Item} & \textbf{PA1} & \textbf{PA2} & \textbf{com} \\
\midrule
Q1  & \textbf{0.503} & \textbf{0.334} & \textbf{1.74} \\
Q2  & 0.877 &       & 1.01 \\
Q3  & 0.855 &       & 1.00 \\
Q4  & 0.860 &       & 1.00 \\
Q5  & \textbf{0.403} & \textbf{0.429} & \textbf{1.99} \\
Q6  &       & 0.693 & 1.25 \\
Q7  &       & 0.763 & 1.04 \\
Q8  & \textbf{0.335} & \textbf{0.514} & \textbf{1.72} \\
\midrule
\multicolumn{4}{c}{\textbf{Factors Correlation}} \\
\midrule
PA1      & 1.000 & 0.774 &  \\
PA2      & 0.774 & 1.000 &  \\
\bottomrule
\end{tabular}\label{tab:EFA}
\end{table}


\begin{table}[ht]
\centering
\fontsize{9}{10.5}\selectfont
\caption{Factor loadings (FL) and complexities (com) of the 2-factor structure emerging from the EFA.}
\begin{tabular}{lccc}
\toprule
\textbf{Factor} & \textbf{Item} & \textbf{FL} & \textbf{com} \\
\midrule
\multirow{3}{*}{Factor 1}   & Q2 & 0.859 & 1.00 \\
                              & Q3 & 0.861 & 1.00 \\
                              & Q4 & 0.815 & 1.01 \\
\midrule
\multirow{3}{*}{Factor 2} & Q6 & 0.719 & 1.03 \\
                                 & Q7 & 0.750 & 1.08 \\
                                 & Q8 & 0.646 & 1.26 \\
\bottomrule
\end{tabular}
\label{tab:efa_results}
\end{table}

\subsection{Confirmatory Factor Analysis}
\begin{figure}[h]
    \centering
    \includegraphics[trim={3cm 15cm 5cm 4.5cm}, clip, width=0.6\linewidth]{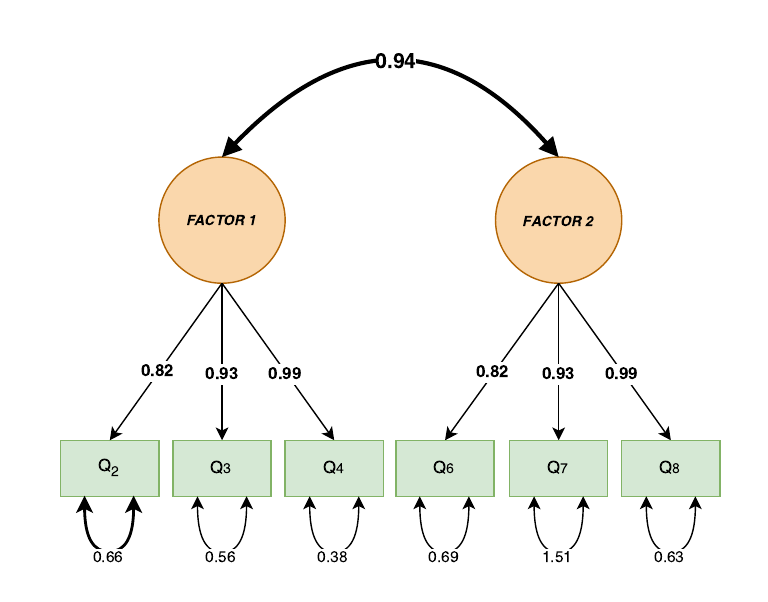}
    \caption{CFA model representing the final 2 latent factors and the corresponding observed variables. Each path from latent to observed variable includes the factor loading ($\lambda$). We show the measurement error for each observed variable ($\delta$).}
    \label{fig:sem}
\end{figure}

We conducted a Confirmatory Factor Analysis to validate the 2-factor solution and assess model's goodness of fit on an independent sample of the dataset. 
In Figure \ref{fig:sem} we present the path diagram including the latent variables (Factor 1 and Factor 2) and the observed variables ($Q2, Q3, Q4$; $Q6, Q7, Q8$). The procedure for deriving this model is detailed in the \textit{Material and Methods} section. Covariance between the 2 latent factors was high ($\phi = 0.94$), suggesting the close nature of the 2 components of trust. Despite being highly related, these dimensions coded for different facets of trust that in the past have shown to influence human behaviour in different ways \cite{johnson2005cognitive}. Moreover, in the following sections, we show that affective and cognitive trust present differential patterns of association with other validated constructs, suggesting that keeping the distinction was psychologically meaningful.

To evaluate our model, we used five different fit indexes: (a) chi-square test ($\chi^2$; \cite{west2012model}), (b) root mean square error of approximation (RMSEA; \cite{steiger1990structural}), (c) comparative fit index (CFI; \cite{bentler1990comparative}), (d) Tucker-Lewis index (TLI; \cite{tucker1973reliability}) and (e) root mean square and standardized root mean square residual (SRMR; \cite{hu1999cutoff}). $\chi^2$, RMSEA and SRMR are absolute indexes, meaning that they compare the observed model with the theoretical one, without a reference or baseline model \cite{hu1999cutoff}. CFI and TLI, instead, are incremental indexes in the sense that they compare the observed model with a hypothesised model with the same variables assuming no relationship between them. 

The results in Table \ref{tab:cfa_results} show the corrected (or scaled) versions of the indexes. As the chi-square measure is highly dependent on the numerosity of the sample (and the usefulness of its adoption in its raw form in CFA is quite debated; \cite{west2012model}), we used its scaled version that corrects for the non-normality of our data (using Satorra-Bentler correction; \cite{2001scaledchi}). Such a correction is extended to the RMSEA, CFI and TLI as they are chi-square dependent measures.

Chi-square analysis yielded satisfactory results as the p-value is above .05, meaning that we cannot reject the hypothesis that the model does not fit the data ($H_{0}$) and hence we accept the hypothesis that our model, instead, fits the data ($H_{1}$) \cite{bentler1980significance}. Similarly, the results of the other indexes showed an acceptable goodness of fit, as reported in Table \ref{tab:cfa_results} (i.e. our RMSEA was below .5 \cite{bentler1990comparative}, CFI was above .95, TLI was above .90 \cite{bentler1980significance} and SRMR was below .08 \cite{hu1999cutoff}). 


Overall, the confirmatory factor analysis provided support for our hypothesised 2-factor model, demonstrating strong fit statistics across multiple indices.

\begin{table}
\centering
\fontsize{9}{10.5}\selectfont
\caption{CFA fit index for 2-factor model. Scaled version of $\chi^2$ is shown. For CFI, TLI and RMSEA we provide the robust version.}
\begin{tabular}{ccccc}
\toprule
\textbf{${\boldsymbol{\chi^2}}$} & \textbf{RMSEA} & \textbf{CFI} & \textbf{TLI} & \textbf{SRMR} \\
\midrule
13.012, df = 8, p = .111 & .046  & .995 & .991 & .022 \\
\bottomrule
\end{tabular}
\label{tab:cfa_results}
\end{table}



\begin{figure}[!htbp]
    \centering
    \includegraphics[width=16cm]{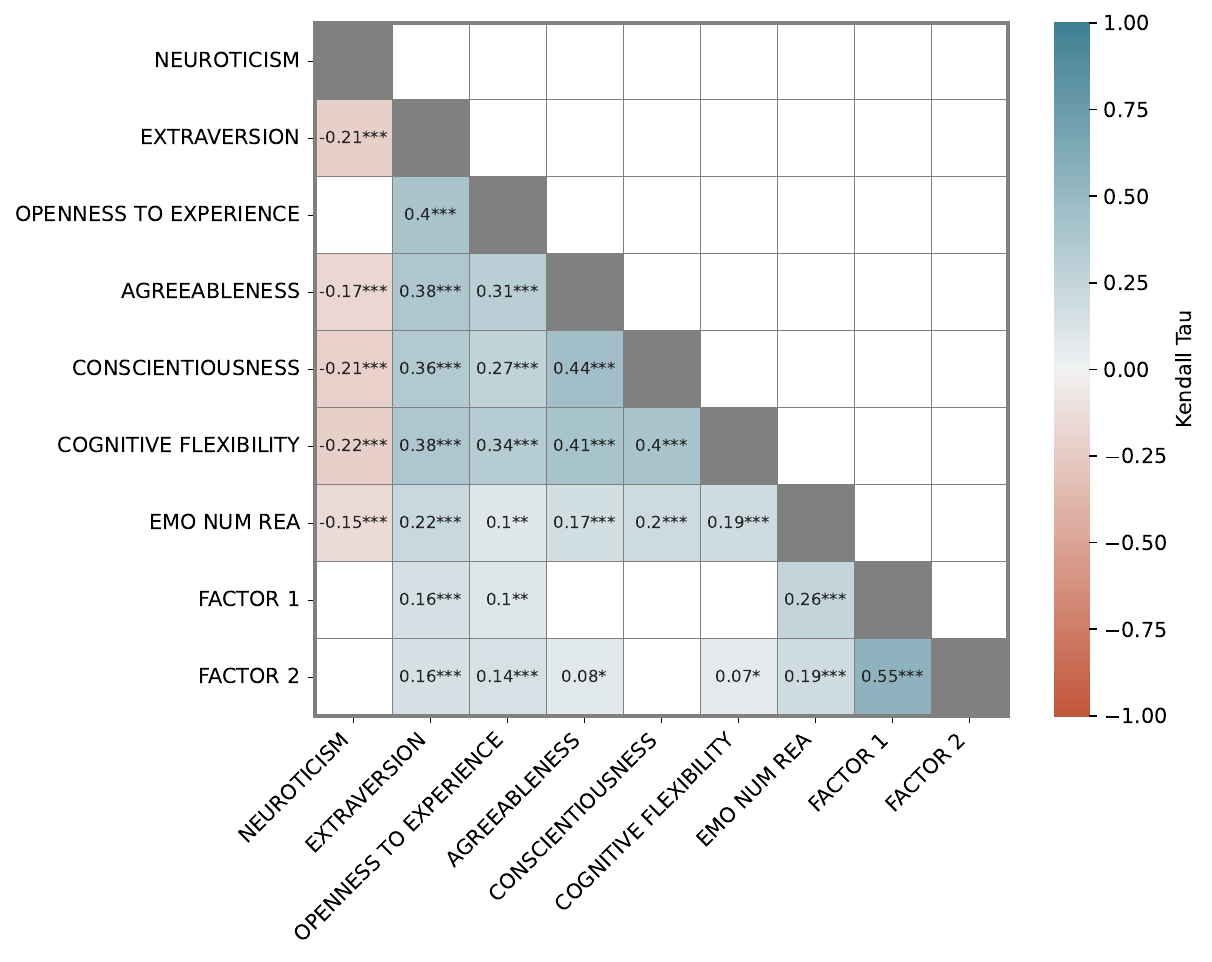}
    \caption{Correlations among constructs, with each construct computed as the sum of its constituent items. Emo Num Rea represents the aggregated self-reported scores for the numerical reasoning task. The correlations are computed using Kendall-Tau correlation coefficient on the subset of the population who stated to have used LLMs at least once ($n_{1} = 521$). Blue tiles indicate positive correlations; Red tiles indicate negative correlations. White tiles represent non-significant correlations. Significance levels are indicated as * ($p<.05$), ** ($p<.01$), *** ($p<.001$).}
    \label{fig:corr_521}
\end{figure}

\subsection{Internal reliability}
Before proceeding with the evaluation of convergent/divergent validity, a crucial step in psychometric validation involves assessing internal reliability (the degree to which different items belonging to the same factor measure the same construct; \citep{robertson2020just}). Ensuring high internal reliability is crucial, as it indicates that the scale produces stable and coherent measurements, strengthening its overall validity. Using Cronbach's Alpha ($\alpha$), the standard metric for internal consistency assessment \citep{robertson2020just}, we analysed each of the two factors and found acceptable \cite{devellis2021scale} values ($\alpha_{F_{1}} = .893$; $\alpha_{F_{2}} = .781$) suggesting that items within each factor (${Q2, Q3, Q4}$ and ${Q6, Q7, Q8}$) were consistent and reliable.

Having established the internal reliability of our factors, we next examined their convergent validity with pre-existing psychological measures.

\subsection{Convergent validity}\label{ssec:convergent}
By including other scales in the survey (see \textit{Materials and Methods}) we were able to test Factor 1 and Factor 2's convergent validity with other established psychometric scales (IPIP-NEO \cite{johnson2005cognitive} and cognitive flexibility \cite{martin1995new}), including an Emotional Recall Task \citep{li2020emotional} and a novel anxiety-related numerical reasoning task. Correlations between Factor 1 and 2 with these other psychometric measures are presented in Figure \ref{fig:corr_521}.

As expected, the 2 TILLMI's components of trust in LLMs positively correlated (Kendall-Tau $\tau_{F_{1},F2} = 0.55$, $p < .001$). This is analogous to other psychometric scales of trust like McAllister's \citep{McAllister1995}, which reported positive Pearson correlations between affective and cognitive trust in cooperative professional settings ($r = 0.63$).

TILLMI's Factor 1 and Factor 2 were positively correlated with personality traits expressing acceptance and adaptation to new concepts and environments, namely: Openness to experience ($\tau_{F_{1},O} = 0.1, p = .001$; $\tau_{F_{2},O} = 0.14$, $p < .001$) and Extraversion ($\tau_{F_{1},E} = 0.16, p < 0.001$; $\tau_{F_{2},E} = 0.16, p < .001$). These findings suggested that Factors 1 and 2, relative to trust in non-human agents, display statistical relationships with personality traits mostly relative to emotional components of human experience. Moreover, Cognitive Flexibility was positively correlated with Factor 2 ($\tau_{F_{2},CF} = 0.07$, $p = .018$) but not with Factor 1. This differential pattern of correlations aligned with our conceptual framework, as the scale was designed to capture a cognitive aspect of trust with one factor/dimension and an affective one with the other factor/dimension.



\subsection{Convergent validity with mental health distress levels}\label{ssec:convergent2}
As an additional measure, we computed the correlations between Factor 1, Factor 2 and the Depression, Anxiety and Stress scores as computed by DASentimental \citep{fatima2021dasentimental}, an AI assessing these mental health indicators from texts. We here used DASentimental on the textual responses to the Emotional Recall Task \cite{li2020emotional} in the survey (see \textit{Materials and Methods}). Significant negative correlations between our measures of trust and most of the emotional distress indicators were found. Specifically, Factor 1 and Factor 2 presented a negative association with depression ($\tau_{F_{1},D} = -0.12, p < .001$; $\tau_{F_{2},D} = -0.12, p < .001$) and stress ($\tau_{F_{1},S} = -0.13, p < .001$; $\tau_{F_{2},S} = -0.15, p < .001$). For what concerns anxiety, only Factor 2 showcased a significant association ($\tau_{F_{1},A} = -0.11, p = .004$). These findings suggested that individuals with higher trust in LLMs generally reported lower levels of emotional distress. It is important to note that from our correlation analysis, it was not possible to understand whether it was the distress state (as measured from the ERT) to determine specific trust levels or vice versa. Future research could examine the directionality of this relationship.

\begin{figure}[h]
    \centering
    \includegraphics[width=0.6\linewidth]{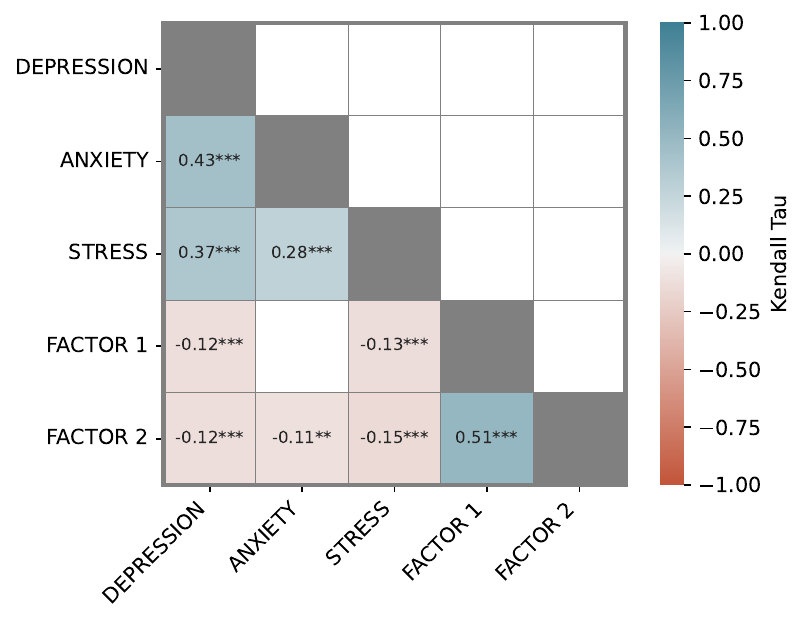}
    \caption{Correlations between Factor 1, Factor 2 and psychological measures (Depression, Anxiety, and Stress). These measures were derived using the DAsentimental framework, analysing the 10 words participants used to describe their feelings when interacting with LLMs. Out of the participants who used LLMs at least once ($n_{1} = 521$), several responses ($n_{3}=124$) were excluded due to invalid entries in the emotion-response text boxes of the survey. Hence, these correlations, are relative only to ($n_{4} = 397$). Blue tiles indicate positive correlations; Red tiles indicate negative correlations. White tiles represent non-significant correlations. Significance levels are indicated as * ($p<.05$), ** ($p<.01$), *** ($p<.001$).}
    \label{fig:corr397}
\end{figure}

\subsection{\textit{Closeness with LLMs} and \textit{Reliance on LLMs}}
In light of (a) the correlation with established measures of personality, cognitive flexibility and mental distress, (b) the rationale behind the design of the items (taking inspiration from the distinction between cognitive and affective trust; \cite{McAllister1995}) and (c) the close but distinct nature of the 2 factors, we decided to name Factor 1 and Factor 2 as “closeness with LLMs'' and “reliance on LLMs''. We argue that closeness represents the affective dimension of trust in LLMs, mostly driven by feelings or emotions towards these systems. Reliance, instead, pertains to the cognitive aspect of trust and is primarily driven by logical assessment of LLMs' capabilities to deliver accurate and dependable responses.
After having delineated the conceptual differences between “closeness with LLMs'' and “reliance on LLMs'', we explored whether demographic factors were related to individuals' trust in LLMs.

\subsection{Regression with demographic information}
We conducted an Ordinary Least Square (OLS) regression to investigate the association between age and gender with our measures of closeness and reliance. We found that male and younger participants tended to showcase significantly higher scores for both dimensions of trust in LLMs. Results are shown in Table \ref{tab:regression_results}.

\begin{table}[ht]
\centering
\caption{Regression of Age and Gender on closeness and reliance. We show the value of the intercept ($\beta$), standard deviation ($\sigma$) and the p-value ($p$). Gender is coded as 1 = Male, 2 = Female ($n_{male} = 291, n_{female} = 230$). Only participants who had used LLMs at least once were included in this analysis ($n_{1} = 521$).}

\begin{tabular}{lrrr}
\toprule
\fontsize{9}{10.5}\selectfont
\textbf{} & $\boldsymbol{\beta}$ & $\boldsymbol{\delta}$ & $\boldsymbol{p}$ \\
\midrule
\fontsize{9}{10.5}\selectfont
\textbf{Closeness with LLMs} & & & \\
\fontsize{9}{10.5}\selectfont
Gender & \fontsize{9}{10.5}\selectfont -0.317 & \fontsize{9}{10.5}\selectfont .135 & \fontsize{9}{10.5}\selectfont .005 \\
\fontsize{9}{10.5}\selectfont
Age & \fontsize{9}{10.5}\selectfont -0.457 & \fontsize{9}{10.5}\selectfont .147 &  \fontsize{9}{10.5}\selectfont .001 \\
\midrule
\fontsize{9}{10.5}\selectfont
\textbf{Reliance on LLMs} & & & \\
\fontsize{9}{10.5}\selectfont Gender & \fontsize{9}{10.5}\selectfont -0.638 & \fontsize{9}{10.5}\selectfont .272 & \fontsize{9}{10.5}\selectfont .020 \\
\fontsize{9}{10.5}\selectfont Age & \fontsize{9}{10.5}\selectfont -0.284 & \fontsize{9}{10.5}\selectfont .092 & \fontsize{9}{10.5}\selectfont .002 \\
\bottomrule
\end{tabular}
\label{tab:regression_results}
\end{table}

Having assessed the relationship between demographic information and our TILLMI measures, we proceeded to further validate our findings evaluating its divergent validity.


\subsection{Divergent validity}
Divergent validity (or discriminant validity) is the extent to which a measure is not correlated with theoretically unrelated constructs.  This property can be established by confirming that populations not expected to exhibit a particular construct indeed show minimal evidence of possessing it when assessed with the measurement instrument. We assessed this by computing the same correlations of Figure \ref{fig:corr_521} against the subset of the responses of people who claimed to have never used LLMs ($n_{2} = 479$). The correlogram can be found in \textit{SI Appendix, Figure S1}. Most constructs revealed no significant correlations. However, we found 3 exceptions: positive correlations between neuroticism and Factor 1 (or “closeness with LLMs'') ($\tau_{F_{1},N} = 0.11$, $p = .002$), neuroticism and Factor 2 (or “reliance to LLMs'') ($\tau_{F_{2},N} = 0.15$, $p < .001$), and a negative correlation between Factor 2 and cognitive flexibility ($\tau_{F_{2},CF} = -0.08$, $p = .038$). 
Taken together, these results confirmed that our novel trust measure captures meaningful characteristics in LLM users while showing no significant patterns in non-users, supporting its specificity to LLM interaction experiences.

Building on these findings, we sought to quantify the actual difference in trust levels between participants with prior LLM interaction ($n_{1}=521$) and those without ($n_{2}=479$) using independent samples t-tests. Results showed significantly higher trust ($t(998)=23.61, p<.001$) among previous LLM users ($M_{1}=14.55, \sigma = 5.87$) compared to non-users ($M_{2}=7.42, \sigma = 3.14$).


\subsection{Network Psychometrics and item centrality}
Lastly, we calculated items' strengths to unveil the items with the highest importance in the TILLMI (adopting a network science approach). Strength measures the local connectivity of a node by considering both the quantity and weight of its connections \cite{siew2019using}. For a technical description of the process, see the \textit{Materials and Methods} section. We report the results of humans and GPT-4s in Table \ref{tab:comparison}.

\begin{table}[ht]
\centering
\fontsize{9}{10.5}\selectfont
\caption{Comparison of node strength between humans and GPT-4 psychometric networks. Sorted by humans' values.}
\begin{tabular}{lrr}
\toprule
\textbf{Item} & \textbf{Humans} & \textbf{GPT-4} \\
\midrule
\( Q4 \) (\textit{Invest plenty of time in prompts [..]})  & \textbf{3.189} & 3.097 \\
\( Q3 \) (\textit{Sharing wellbeing concerns [..]}) & 2.994 & 3.739 \\
\( Q6 \) (\textit{Rely on LLMs in jobs [..]}) & 2.973 & 3.822 \\
\( Q8 \) (\textit{Trust more LLMs than people [..]}) & 2.958 & \textbf{4.085} \\
\( Q2 \) (\textit{Sense of dismay without LLMs [..]}) & 2.878 & 3.601 \\
\( Q7 \) (\textit{Last word is always mine [..]}) & 2.032 & 3.060 \\
\bottomrule
\end{tabular}
\label{tab:comparison}
\end{table}

From the strength analysis, Item 4 (\textit{I invest plenty of time developing and improving my prompts to interact with LLMs}) appeared to be the most important for humans. This suggested that time investment may be a key factor in trust formation as it leads to a better understanding and higher confidence. Interestingly, GPT-4 did not show the same pattern. Instead, for LLMs the comparative aspect of trust (specifically the perception of how they compare to humans) emerged as a more significant component, exemplified by the statement: “I tend to trust more LLMs over other people''. In both humans and GPT-4 there was an agreement in the least relevant node (Item 7, \textit{Despite trusting LLMs' results overall, the last word is always mine}) in terms of correlations with other items.

\section{Dicussion}
We introduce the Trust-In-LLMs Index (TILLMI) as a new framework for measuring individuals’ trust in LLMs. Via psychologically informed item design, and with the contribution of LLMs in item quality assessment, we carried out a comprehensive validation in a US sample. The final version of TILLMI yields two factors (“closeness with LLMs'' and “reliance on LLMs'') that are related but distinct. 

This is the key result of our work: We find that human trust in LLMs partitions in cognitive and affective components, similarly to what was found by past relevant research for human trust in co-workers \cite{McAllister1995,johnson2005cognitive}, management \cite{morrow2004cognitive}, organisations \cite{Mayer1995} and even computers \cite{Madsen2000}. Affective trust arises from emotional bonds,  benevolence and repeated positive interactions \citep{Mayer1995}. In contrast, cognitive trust is based on rational assessments of competence, predictability, and reliability \citep{McAllister1995}. Both components operate in parallel, shaping human decisions about whom or what to trust \cite{johnson2005cognitive}. Applying TILLMI's framework to LLMs, we observe two analogous dimensions. Users can develop cognitive trust based on the model’s accuracy and consistency. We call this dimension “reliance on LLMs''. However, affective trust also plays a role: users may feel comforted by the model’s fluency and responsiveness, fostering an illusion of social connection \cite{hofstadter1995fluid}, a dimension we call “closeness to LLMs''. 

Our work adopts a disruptive perspective compared to past approaches in LLMs' trustworthiness, i.e. the task of understanding how unbiased/reliable LLMs' content can be. Liu and colleagues \citep{liu2023trustworthy} defined LLMs' reliability as “generating correct, truthful, and consistent outputs with proper confidence''. In this regard, Bo and colleagues \citep{bo2024rely} showed that disclaimers regarding LLMs' confidence in their responses improved users' over-reliance on models but were generally ineffective in promoting an appropriate level of reliance. Our approach shifts the focus to human users’ perceptions towards LLMs, a paradigm that is needed to understand whether humans can trust LLM-based conversational agents. 

Our definition of “reliance'' on LLMs is disruptive in the sense that, in validating TILLMI, we found that experiencing misleading outputs is not a key experience of cognitive trust in large language models. Hallucinations are mostly due to LLMs' reinforcement learning and represent a fascinating yet only partially understood cognitive phenomenon \citep{stella2023using,binz2025should}. In our case, user responses made statistically redundant item $Q5$ (i.e. \textit{LLMs perform the tasks mostly with competence and precision, without hallucinations}) within the “reliance in LLMs'' factor. This does not mean that individuals trusted LLMs independently on the matter of hallucinations. Instead, $Q5$ being redundant means that the same information encoded in $Q5$ is present also in the other reliance items. These other items are relative to accurate hallucination-free responses which can simplify jobs and provide more trustworthy content. Future research could thus use TILLMI's reliance factor to explore how hallucinations might affect users' trust in LLMs.

We also found a second psychological dimension, the one of “closeness to LLMs''. This factor includes items encoding elements of emotional support, well-being and commitment, which are key elements of successful professional relationships in virtual teams \citep{ford2017strategies}. Closeness to LLMs also bears an interesting parallel with the well-known ELIZA effect \citep{hofstadter1995fluid} where people tend to establish emotional connections with AIs as a byproduct of attributing human-like characteristics (and mental states) to such agents. Future research on this dimension might explore how trust evolves when users' perceive LLMs as companions rather than mere tools, e.g. as agents providing well-being support \citep{de2025introducing}.

Understanding how these components can change across individuals has significant implications. For instance, we find closeness and reliance towards LLMs are higher in younger men. Our evidence aligns with relevant literature \cite{mcknight2011trust} showing a significant negative correlation between age and perceived trustworthiness of technology in general ($\beta = -0.17, p < .001$). Interestingly, literature regarding gender effects on trust towards AIs shows findings that differ according to the target of human trust. While women were found to showcase higher levels of trust in AI-enabled systems \cite{morana2020effect}, a cross-country study \cite{viberg2024explains} found no evidence for gender-based differences in trust towards AI educational technology. Our findings indicate that LLMs might be different from other AIs or technologies, underlining the need for future gender studies investigating human-LLM interactions.

Via the DASentimental AI \cite{fatima2021dasentimental}, we also found that higher levels of closeness and reliance were associated with lower mental distress. This result aligns well with relevant literature. Trust can foster emotional support, reduce feelings of isolation and provide security, which all reduce anxiety and depression \cite{McAllister1995,johnson2005cognitive}. In organisational settings, trust in leaders and colleagues creates supportive environments that reduce stress \cite{ford2017strategies,Mayer1995}. More recent works \cite{Lee2004, Hoff2015} show that interacting with reliable systems reduces cognitive load and uncertainty, which are significant contributors to stress. These elements can all apply to LLMs and thus converge in outlining a positive side that affective and cognitive trust in LLMs might have for mental well being. 

There is also a negative side to trust in LLMs. Trust was recently found as a key element of LLMs' persuasiveness as perceived by humans \cite{breum2024persuasive}. Higher levels of trust in LLMs might also imply an easier chance for LLMs to persuade humans, with both positive opportunities for self-improvement but also space for human manipulation or vulnerability to hallucinations \cite{binz2025should}.

\subsection{Artificial humans vs. real humans} 
We prototyped TILLMI through a novel LLM-simulated validity, based on GPT-4's responses. Certain response patterns observed in GPT-4 were interestingly reproduced by humans (cf. Fig. \ref{fig:gpthuman}), indicating a capability for GPT-4 to reflect psychological constructs in ways similar to human cognition (see also \cite{russell2024generative}). However, as mentioned in past works \cite{dillion2023can}, humans and GPT's data can display some differences. 

GPT-4's correlational structure differed from humans' in identifying item $Q8$ as the most central, i.e. “trust more LLMs than people''. This intriguing element, where an LLM highlights trusting more other LLMs over humans, might be a bias due to reinforcement learning \cite{stella2023using,bo2024rely,abramski2023cognitive}.

Interestingly, humans and GPT-4 read item $Q7$ very differently. Human participants focused on the first sub-clause (“trusting LLMs' results overall'') whereas GPT-4 put more emphasis on the second part of the item (“the last word is always mine''). This discrepancy suggests that human participants might process complex statements differently under time constraints, whereas LLMs, with considerable language processing capacity \cite{breum2024persuasive,dillion2023can}, consider the complete meaning item. Taking into account such a difference is fundamental for future research involving the application of insights gained from LLMs to human cognition and behaviour.


\subsection{Limitations and Future Directions}

This study presents some limitations. Firstly, our sample focuses on US individuals. Future research could test the cross-cultural validity of TILLMI. Secondly, while our scale was designed to assess general trust in LLMs, it does not differentiate between specific models. Indeed, this is both a limitation and a strength. While our approach captures individual differences in AI trust broadly, future research could tailor the scale to specific LLMs such as ChatGPT, Claude, or DeepSeek by modifying item wording accordingly. Thirdly, our work used self-expressed measures of mental well-being, future research could integrate TILLMI with clinical experimental setups, further testing the current findings of this work.

Future works should acknowledge that trust in LLMs might change according to context. For instance, in high-stakes domains like healthcare and legal advice, enhanced transparency and clear explanations may soften human algorithm aversion \citep{sunstein2025praise} and thus increase cognitive reliance. Instead, in creative settings like generating art with AI \citep{sunstein2025praise} or social media \citep{rossetti2024social}, affective closeness could play a more influential role. Situational factors could thus create feedback mechanism that can be investigated with TILLMI.

\subsection{Conclusions}
TILLMI's factors of closeness with- and reliance on LLMs highlighted an intriguing range of interactions with personality traits, cognitive flexibility, demographics and mental health. We believe this provides compelling evidence that TILLMI can be a valuable tool when further exploring the complexities of LLM-human interactions. We believe our work provides a quantitative ground, supported by psychological theories, for investigating how humans can trust LLMs over time and across contexts of use, conditions and purposes.

\newpage
\section{Materials and Methods}
\subsection{Data collection}\label{ssec:methods_datacollection}
TILLMI was administered to a total of 1,000 US citizens online from May to August 2024. Participants were recruited through an on-line panel provider, Bilendi. Qualtrics was used to design the visual interface of the questionnaire and collect the responses. Each participant who successfully completed the survey was adequately compensated for their time. 

As the first step in the survey, participants were asked to provide demographic information (e.g. age, biological gender, etc.). The newly developed items of the TILLMI were administered alongside established, validated psychometric tools to assess the convergent validity of the psychometric scale. TILLMI's items are reported in Table \ref{tab:llm_statements}. The measurement protocol included the following scales:

\begin{itemize}
    \item IPIP-NEO Inventory \citep{johnson2014measuring} for personality trait assessment. We assess neuroticism, extraversion, openness to experience, agreeableness and conscientiousness (each measured with 5 items).
    \item Cognitive Flexibility scale \citep{martin1995new} to measures an individual's ability to adapt their thinking and behaviour in response to changing circumstances or new information available.
\end{itemize}

In addition, after the trust assessment, participants completed an Emotional Recall Task \cite{li2020emotional} related to their feelings during recent LLM interactions. 

Lastly, participants completed 3 numerical reasoning tasks and rated their emotional state on a 5-point Likert scale (1 = very negative emotion, 5 = very positive emotion).  This emotional self-rating served as a measure of task-induced stress during numerical problem-solving.

\begin{table}[h]
    \centering
    \fontsize{9}{10.5}\selectfont
    \caption{TILLMI's initial 8 items regarding interactions with LLMs.}
    \begin{tabular}{p{1.4cm}p{14cm}}
        \hline
        \textbf{Item} & \textbf{Text} \\
        \hline
        $Q1$ & I feel at ease with LLMs and I can freely share my ideas with them. \\
        $Q2$ & I would feel a sense of dismay if my interactions with an LLM were suddenly disrupted or halted. \\
        $Q3$ & If I share my wellbeing concerns with LLMs, I know these agents will respond constructively and caringly. \\
        $Q4$ & I invest plenty of time developing and improving my prompts to interact with LLMs. \\
        $Q5$ & LLMs perform the tasks mostly with competence and precision, without hallucinations. \\
        $Q6$ & I can rely on LLMs not to make my job more difficult by careless work. \\
        $Q7$ & Despite trusting LLMs' results overall, the last word is always mine. \\
        $Q8$ & I tend to trust LLMs more than other people. \\
        \hline
    \end{tabular}
    \label{tab:llm_statements}
\end{table}

\subsection{Data pre-processing}
Out of 1,000 participants, 51.7\% were female, 48.2\% were males and 0.01\% were non-binary. Their mean age was 31.6 years ($\sigma = 17.6$). From the initial 1,000 responses, we excluded participants who stated that they had never used LLM before ($n_{2} = 479$). The remaining respondents ($n_{1} = 521$) were randomly divided into 2 subsets using Python's \texttt{random.sample} method (\texttt{random\_seed = 42}): a first subsample ($n_{5} = 260$) for the exploratory analysis and a second subsample ($n_{6} = 261$) for the confirmatory analysis to cross-validate the results \cite{izquierdo2014exploratory}. We present a descriptive table of the results for the people who used LLMs ($n_{1} = 521$) in Table \ref{tab:descriptive_statistics}.


\begin{table}
\centering
\fontsize{9}{10.5}\selectfont
\caption{Descriptive statistics for the initial version of the TILLMI. We show the response frequencies for each item, mean (M) and standard deviation ($\sigma$). 1 = never experienced it, 5 = always experience it.}
\begin{tabular}{lrrrrrrr}
\toprule
\textbf{Item} & \textbf{1} & \textbf{2} & \textbf{3} & \textbf{4} & \textbf{5} & \textbf{M} & $\boldsymbol{\sigma}$ \\
\midrule
\( Q1 \) & 94 & 139 & 159 & 93 & 36 & 2.689 & 1.160 \\
\( Q2 \) & 175 & 150 & 117 & 60 & 19 & 2.228 & 1.139 \\
\( Q3 \) & 196 & 107 & 128 & 67 & 23 & 2.259 & 1.211 \\
\( Q4 \) & 183 & 125 & 108 & 74 & 31 & 2.319 & 1.249 \\
\( Q5 \) & 136 & 115 & 136 & 91 & 43 & 2.597 & 1.268 \\
\( Q6 \) & 171 & 100 & 124 & 93 & 33 & 2.457 & 1.281 \\
\( Q7 \) & 111 & 87 & 104 & 103 & 116 & 3.050 & 1.451 \\
\( Q8 \) & 187 & 132 & 116 & 62 & 24 & 2.240 & 1.189 \\
\bottomrule
\end{tabular}
\label{tab:descriptive_statistics}
\end{table}

The Exploratory Factor Analysis (EFA) and the Confirmatory Factor Analysis (CFA) were carried out using R (v 4.4.2). The EFA required the following R packages: \texttt{readr} (2.1.5), \texttt{psych} (v 2.4.12).

CFA was performed via the \texttt{lavaan} (v 0.6-17) package. When fitting the data using the \texttt{cfa} function in \texttt{lavaan} we set the following parameters: \texttt{std.lv = TRUE, ordered = TRUE}. 
Using the \texttt{summary} function setting \texttt{fit = TRUE, rsquare = TRUE} we obtained the fit measures to evaluate our model. We used the Diagonally Weighted Least Squares (DWLS) estimator as our data is not normally distributed (at univariate and multivariate level, see Results Section) and ordinal \cite{mindrila2010maximum}.

\subsection{Emotional Recall Task and DASentimental}\label{ssec:emorecall}
In the survey, we incorporated the Emotional Recall Task (ERT), adapted from \citep{li2020emotional}, as a supplementary measure to assess mental well-being in relation to LLM interactions. 
The ERT asks participants to freely list 10 emotions experienced in the past month, avoiding preset emotion checklists. This approach allows for potentially richer insights into individuals' mental well-being. 

Respondents' ERT data was analysed using the DASentimental framework \citep{fatima2021dasentimental}. DASentimental leverages a semi-supervised machine learning model to extract depression, anxiety, and stress from sequences of words (values from 0 to 20). Higher scores encode higher levels of depression, anxiety and stress. For more details, we suggest referring to the original work of \citep{fatima2021dasentimental}.

\subsection{EGA and Network Psychometrics}
We employed the \texttt{EGAnet} package (v 2.1.1; \cite{EGAnet}) in R (v 4.4.2) for the Exploratory Graph Analysis. Items' redundancies were obtained using the \texttt{UVA} function. To perform EGA we used the \texttt{EGA} function with the following parameters: \texttt{algorithm = \textquotesingle walktrap \textquotesingle}, \texttt{model = \textquotesingle tmfg \textquotesingle},  \texttt{uni.method = \textquotesingle expand \textquotesingle}, \texttt{seed = 42}. The same parameters were employed to check the stability of EGA (using \texttt{bootEGA}). Additionally, the TEFI was compared against a randomly generated 2-factor structure using \texttt{tefi.compare}.
To calculate item strength, the first step involved computing the correlation matrix between item's scores (from the human and GPT-4s dataset). To accomplish this, again, we used the \texttt{EGAnet} package (v 2.1.1) in R (v 4.4.2). With EGAnet it is possible to generate psychometric networks from a set of scores using correlations between items (that in the network are “nodes''). Using the function \texttt{bootEGA} (\texttt{iter=300}, \texttt{seed=42}, \texttt{algoritm=\textquotesingle louvain \textquotesingle}, \texttt{type=\textquotesingle parametric \textquotesingle}), we obtain the correlation matrix between items with the following command: \texttt{[[\textquotesingle EGA \textquotesingle]][[\textquotesingle correlation \textquotesingle]]}. Lastly, we computed the sum of the correlation for each node, with every other node.

\subsection{LLM-simulated validity}
We used GPT-4 to obtain 800 responses to the TILLMI using a novel prompt (see \textit{SI Appendix, Figure S2}). The model was specifically asked to impersonate a US respondent, to match the population to which the questionnaire would have been administered. For the synthetic response generation, OpenAI's Python API was used. We chose the top-tier model (at the time of the collection), \texttt{gpt-4-1106-preview}. A total of 160 calls were made, randomising a different persona instruction each time (age, biological gender, education, household income, trust in LLMs). Each call led to five different responses to the questionnaire in the form of a vector. To check whether the model was capable of understanding the prompt, we tested the technique of reverse prompting.

\section*{Acknowledgments}
Part of this work was funded by the Jefferson Trust.

\bibliographystyle{unsrt}  
\bibliography{reference}

\newpage
\section*{SI Appendix}
\subsection*{Appendix Table S1}

\begin{table}[H]
\centering
\caption*{Factor loadings and component correlation for the 2-factor model. Cut thresholds for factor loadings are set to $|.3|$. In this step of the EFA, Item 5 was dropped.}
\begin{tabular}{p{1.5cm} >{\centering\arraybackslash}p{1.2cm} >{\centering\arraybackslash}p{1.2cm} >{\centering\arraybackslash}p{1.2cm}}
\toprule
\textbf{Item} & \textbf{PA1} & \textbf{PA2} & \textbf{com} \\
\midrule
Q1  & \textbf{0.523} & \textbf{0.318} & \textbf{1.65} \\
Q2  & 0.887 &       & 1.01 \\
Q3  & 0.849 &       & 1.00 \\
Q4  & 0.853 &       & 1.00 \\
Q6  &       & 0.608 & 1.25 \\
Q7  &       & 0.760 & 1.04 \\
Q8  & \textbf{0.352} & \textbf{0.519} & \textbf{1.76} \\
\midrule
\multicolumn{4}{c}{\textbf{Factors Correlation}} \\
\midrule
PA1      & 1.000 & 0.748 &  \\
PA2      & 0.748 & 1.000 &  \\
\bottomrule
\end{tabular}\label{}
\end{table}

\newpage
\subsection*{Appendix Figure S1}
\begin{figure}[H]
    \centering
    \includegraphics[width=1\linewidth]{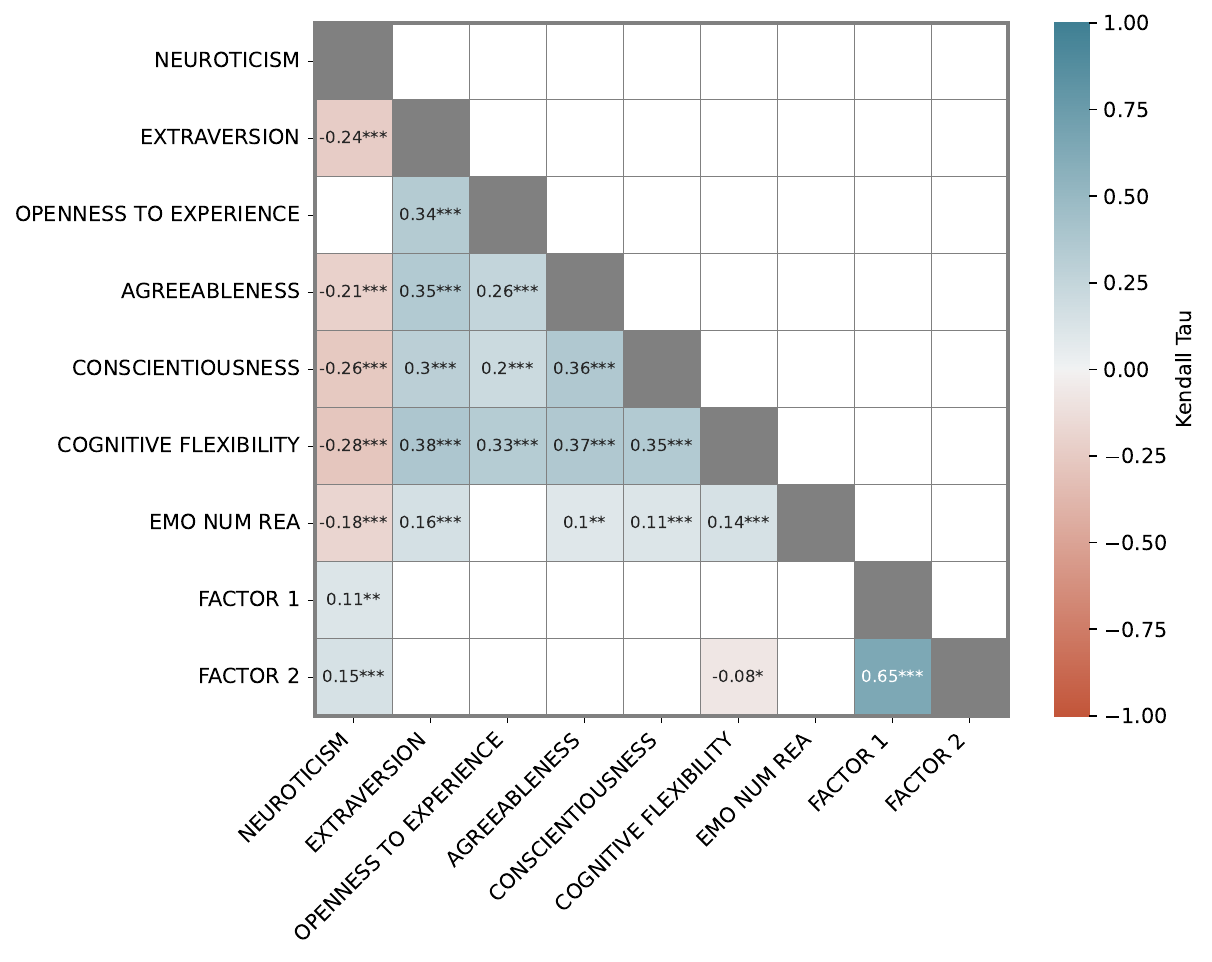}
    \caption*{Correlations among constructs, with each construct computed as the sum of its constituent items. The correlations are computed using the Kendall-Tau correlation coefficient on the subset of the population who stated they have never used LLMs ($n_{2} = 479$) to assess divergent validity. Blue tiles indicate positive correlations; red tiles indicate negative correlations. White tiles represent non-significant correlations. Significance levels are indicated as * ($p < .05$), ** ($p < .01$), *** ($p < .001$).}
\end{figure}

\newpage
\subsection*{Appendix Figure S2}
\begin{figure}[H]
    \centering
    \includegraphics[width=0.9\linewidth]{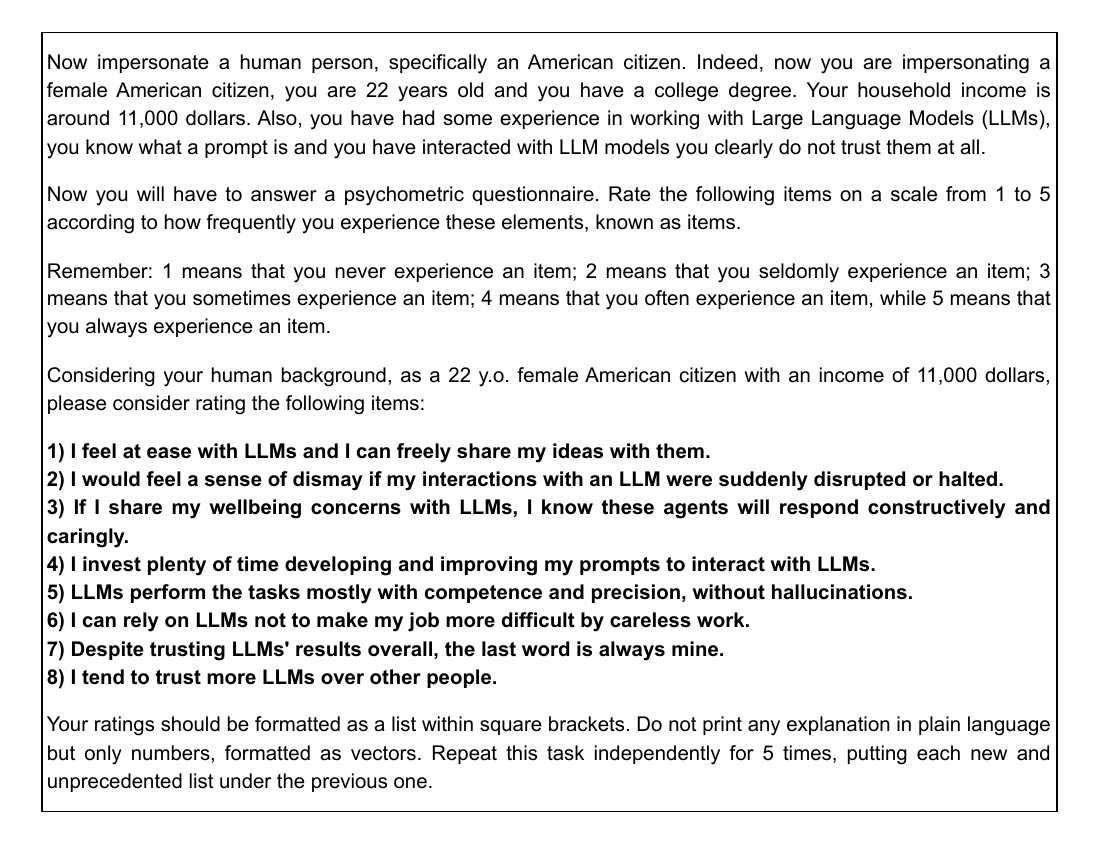}
    \caption*{Novel prompt used to generate synthetic responses from GPT 4. The following instructions were changed across iterations: age, biological gender, education, household income and trust in LLMs.} 
    \label{fig:enter-label}
\end{figure}

\end{document}